\documentstyle[emulateapj]{article}

\catcode`\@=11 
\def\@versim#1#2{\vcenter{\offinterlineskip
        \ialign{$\m@th#1\hfil##\hfil$\crcr#2\crcr\sim\crcr } }}

\newcommand{\beq}{\begin{equation}}
\newcommand{\eeq}{\end{equation}}
\def\lsim{\mathrel{\mathpalette\@versim<}}
\def\gsim{\mathrel{\mathpalette\@versim>}}
\def\mpy{M_\odot \ {\rm yr^{-1}}}
\def\rt{r_{\rm tr}}
\def\ro{r_{\rm o}}
\def\rc{r_{\rm crit}}
\def\ti{t_{\rm inf}}
\def\tc{t_{\rm cool}}
\def\md{\dot M_o}
\def\cxo{{\it CXO}}
\def\gcc{{\rm g \ cm^{-3}}}
\def\kms{{\rm km \ s^{-1}}}
\begin{document}

\title{The Cooling Flow to Accretion Flow Transition} 

\author{Eliot Quataert and Ramesh Narayan} \affil{Harvard-Smithsonian
  Center for Astrophysics, 60 Garden St., Cambridge, MA 02138;
  equataert, rnarayan, @cfa.harvard.edu}

\medskip

\begin{abstract}
  
  Cooling flows in galaxy clusters and isolated elliptical galaxies
  are a source of mass for fueling accretion onto a central
  supermassive black hole.  We calculate the dynamics of accreting
  matter in the combined gravitational potential of a host galaxy and
  a central black hole assuming a steady state, spherically symmetric
  flow (i.e., no angular momentum).  The global dynamics depends
  primarily on the accretion rate.  For large accretion rates, no
  simple, smooth transition between a cooling flow and an accretion
  flow is possible; the gas cools towards zero temperature just inside
  its sonic radius, which lies well outside the region where the
  gravitational influence of the central black hole is important.  For
  accretion rates below a critical value, however, the accreting gas
  evolves smoothly from a radiatively driven cooling flow at large
  radii to a nearly adiabatic (Bondi) flow at small radii.  We argue
  that this is the relevant parameter regime for most observed cooling
  flows.  The transition from the cooling flow to the accretion flow
  should be observable in M87 with the {\it Chandra X-ray
  Observatory}.


\ 

\noindent {\em Subject Headings:} galaxies: cooling flows -- accretion, accretion disks -- galaxies:  individual (M87)

\end{abstract}

\section{Introduction}

X-ray observations of clusters of galaxies and isolated elliptical
galaxies indicate the presence of significant quantities of hot gas
whose radiative cooling time is much less than a Hubble time (see,
e.g., Fabian 1994 for a review).  This is generally interpreted as
evidence for a cooling flow, i.e., the radiatively driven accretion of
gas onto a central cluster galaxy or towards the center of the host
galaxy (e.g., Cowie \& Binney 1977; Fabian \& Nulsen 1977; Mathews \&
Bregman 1978).  The inferred accretion rates range from $\sim 0.1
\mpy$ to $\sim 10^3 \mpy$ and are often inferred to decrease
substantially with decreasing radius, roughly as $\dot M \propto r$
(e.g., Stewart et al.  1984).  The decreasing $\dot M$ is probably
evidence for an inhomogeneous flow, in which some of the gas (that
which is denser) cools out of the hot flow at large radii.



It has long been recognized that a cooling flow may be an important
source of fuel for accretion onto a central supermassive black hole in
the host galaxy (e.g., Bailey 1980, Nulsen, Stewart, \& Fabian 1984).
The goal of this paper is to calculate in some detail the dynamics of
accreting matter in the central regions of a galaxy which harbors both
a cooling flow and a supermassive black hole.  We restrict ourselves
to spherically symmetric flows.


Our model and its assumptions are discussed in the next section (\S2).
In \S3 we present solutions to our model problem and explain the
physics which governs the dynamical transition from a cooling flow to
an accretion flow.  Finally, \S4 sums up and discusses the
observational implications of our work.

\section{The Model}

We consider a simplified model which we believe captures much --
although not all (\S4) -- of the interesting physics of the transition
from a cooling flow to an accretion flow.  The primary assumptions of
our model are: (1) steady state and (2) spherical symmetry.  Aside
from adiabatic compression, we assume negligible heating of the gas,
and we approximate its cooling as due to bremsstrahlung emission.

Cooling flows are inferred not to conserve mass.  Theoretically, this
can be understood by noting that the cooling time of the gas is
comparable to the inflow time.  Overdense ``blobs'' can therefore
condense out of the flow before accreting.  We model this by including
a mass drop out term in the continuity equation (cf White \& Sarazin
1987):

\beq { d \dot M \over d r} = q {\dot M \over r } {\ti \over \tc},  \ \  \ \ 
\dot M = - 4 \pi r^2 \rho v.
\label{mdot} \eeq
Momentum and energy conservation are given by

\beq v {d v \over d r} = - {d \phi \over d r} - {1 \over \rho} {d \rho
  c^2_s \over d r} \label{rad} \eeq

and

\beq
\rho v T {d s \over d r} = {\rho v \over \gamma - 1} {d c^2_s \over dr} - c^2_s v {d \rho \over d r} = - B \rho^2 c_s. \label{energy} \eeq

The quantities $\dot M$, $\rho$, $v$, $\phi$, $c_s$, $T$, $s$, and
$\gamma$ represent the mass accretion/inflow rate, the mass density,
the radial velocity (negative inwards), the gravitational potential,
the isothermal sound speed, the temperature, the entropy per unit
mass, and the gas adiabatic index, respectively.  The quantity $B
\approx 5.6 \times 10^{16}$ ergs cm$^2$ g$^{-2}$ is the normalization
for the bremsstrahlung emission.

The efficiency of mass drop out depends on the ratio of the inflow
time of the gas ($\ti \equiv r/|v|$) to the cooling time ($\tc \equiv
c_s/B \rho$).  When $\ti \sim \tc$ (as in a cooling flow), mass drop
out can be important since there is time for overdense regions to
condense out.  When $\ti \ll \tc$, however (as in Bondi's spherical
accretion solution), mass drop out is negligible since the gas
accretes before cooling.  The parameter $q$ in equation (\ref{mdot})
normalizes the efficiency of mass drop out; we expect $q \sim 1$.

We take the gravitational potential to be given by \beq \phi = {-GM
  \over r - r_g} + { \sigma^2 \ln r} \ \ {\rm for} \ \ r > r_b
\label{fg1} \eeq and \beq \phi = {-GM \over r - r_g} -
\sigma^2 \left({ r \over r_b}\right)^{-\beta + 1} \ \ {\rm for} \ \ r
< r_b \label{fg2} \eeq
The left most term in equations (\ref{fg1}) and (\ref{fg2}) is the
gravitational potential due to a black hole of mass $M$ and
Schwarzschild radius $r_g = 2GM/c^2$; the $1/(r-r_g)$ mimics the
effects of General Relativity (Paczy\'nski \& Wiita 1980).

{\it HST} observations of the centers of elliptical galaxies indicate
that those which harbor massive black holes have central surface
brightness ``cores.''  The surface brightness rises steeply with
decreasing radius at large radii, but flattens out in the inner
portions of the galaxy (e.g., Lauer et al.  1995; Faber et al.  1995;
Kormendy \& Richstone 1995).\footnote{Note, however, that the surface
  brightness does not become constant even on the smallest scales
  accessible by {\it HST}.  It still increases with decreasing radius,
  but with a noticeably smaller power law slope.}  Our mass model for
the galaxy in equations (\ref{fg1}) and (\ref{fg2}) is intended to
reflect these observations.  Outside a break radius $r_b$, the galaxy
has a constant velocity dispersion ($\sigma$) while inside that radius
we assume that light traces mass (aside from the black hole) and so
the enclosed mass profile flattens ($\beta < 1$) in accord with the
flattening surface brightness.

Typical observed values of $\beta$ and $r_b$ are $\approx 0.25$ and
$\approx 1$ kpc, respectively (Faber et al. 1995); we use such values
in our numerical results of \S3.1.  For analytical estimates, however,
we take $r_b \rightarrow 0$, i.e., we model the galaxy as having a
constant velocity dispersion everywhere.

The gravitational potential is predominantly that of the galaxy for $r
\gsim \rt$ while it is predominantly that of the black hole for $r
\lsim \rt$, where the transition radius is (for $r_b \rightarrow 0$)
\beq \rt \approx {G M \over \sigma^2} \approx 0.05 \ {\rm kpc} \
\left({M \over 10^9 M_\odot}\right) \left({\sigma \over 300 \ {\rm km
\ s^{-1}}}\right)^{-2} . \label{rt} \eeq For a galaxy with $\sigma
\approx 300$ km s$^{-1}$ in the Virgo cluster, a distance $\approx 20$
Mpc away, $\rt$ corresponds to $\approx 0.5$ arcsec.  This is
comparable to the angular resolution of the {\it Chandra X-ray
Observatory (CXO)}, indicating that the presence of a central black
hole in nearby ellipticals may have observable effects on the cooling
flow X-ray emission (see \S4.1).

\section{Model Solutions}

We are interested in solutions to our model problem which undergo a
subsonic to supersonic transition at a sonic radius $r_s$.  Rewriting
equations (\ref{mdot})-(\ref{energy}) yields \beq v {d v \over d r} =
{N \over D}, \eeq where \beq N = {{-d \phi \over d r} + {2 \gamma
    c^2_s \over r} + {B \rho c_s (\gamma - 1 + \gamma q) \over v}}
\eeq and \beq D = 1 - {\gamma c^2_s \over v^2}.  \eeq Since $D = 0$ at
the sonic point, we must have $N = 0$ for a smooth transition.

We take our outer boundary conditions to be specified values of
$\rho_o$ and $c_o$, the density and sound speed, at an outer radius
$\ro$.  Equations (\ref{mdot})-(\ref{energy}) have two eigenvalues,
taken here to be the sonic radius, $r_s$, and the accretion rate at
the sonic radius, $\dot M(r_s)$.  We find our solutions by shooting
out from $r_s$ and adjusting $\dot M (r_s)$ and $r_s$ to satisfy the
outer boundary conditions.

For most of this paper we scale our models to observations of M87:
$\ro \approx 100$ kpc, $\rho_o \approx 10^{-27}$ g cm$^{-3}$, $\sigma
\approx 300$ km s$^{-1}$, $c_o \approx \sigma$ (e.g., Stewart et al.
1984) and $M \approx 3 \times 10^9 M_\odot$ (Harms et al. 1994; Ford
et al.  1994; Macchetto et al. 1997).  In M87, $\dot M$ is inferred to
decrease from $\approx \, 10 \, \mpy$ at $\approx 70$ kpc to $\lsim \,
1 \, \mpy$ at a few kilo-parsecs (Stewart et al. 1984).  As shown
below, this is reasonably well captured by a $q = 0.6$ model.  

Although strictly an eigenvalue of our problem, an excellent estimate
of the value of $\dot M$ at the outer boundary, $\dot M(\ro) \equiv
\md$, can be made by imposing the requirement that the gas be a
cooling flow at $\ro$.  This requires that the inflow time of the gas
be comparable to the local radiative cooling time which in turn
requires $v(\ro) \equiv v_o \approx - B \rho_o r_o/c_o$ and therefore
(cf Fabian \& Nulsen 1977)
\begin{eqnarray} \label{md} \md &\approx& {4 \pi B \rho^2_o r_o^3 \over
    c_o} \\ \nonumber & \approx & 13 \ \mpy \ \left(\rho_o \over
    10^{-27} \ \gcc \right)^2 \\ \nonumber & \times & 
\left(r_o \over 100 {\rm \ kpc} \right)^3 
\left(c_o \over 300 \ \kms\right)^{-1}.
 \end{eqnarray}
 The primary question is what happens to matter accreting at this rate
 at smaller radii.
 
 In the absence of a central point mass a cooling flow with $\md$
 given by equation (\ref{md}) would have the following structure: \beq
 c_s \approx \sigma, \ \ \dot M \propto r^{3q/(q + 2)},\ \ \rho
 \propto r^{-3/(q + 2)}, \ \ {\&} \ \ v \propto r^{(q - 1)/(q +
   2)}.
 \label{scaling}\eeq This is readily verified by substitution into
 equations (\ref{mdot})-(\ref{energy}) under the assumption that the
 gas is subsonic and that the potential is logarithmic ($M = 0$, $r_b
 \rightarrow 0$).  Note, however, that if $q < 1$ the above scalings
 show that the Mach number of the flow increases as the radius
 decreases.  At $r_o$, the Mach number is
 \begin{eqnarray}
  \label{mach}  {v_o \over c_o} \nonumber &\approx& 0.02 \left(\rho_o \over
     10^{-27}\ \gcc \right) \left(r_o \over 100 {\rm \ kpc} \right)
   \left(c_o \over 300 \ \kms \right)^{-2} \\  & \approx & 0.02
\left({\dot M \over 13 \ \mpy}\right)^{1/2} \left(r_o \over 100 {\rm \ kpc} 
\right)^{-1/2} \\ \nonumber & \times & 
\left(c_o \over 300 \ \kms \right)^{-3/2}.\end{eqnarray}
   Therefore, there is a critical radius at which the flow would make
   a sonic transition: \beq r \equiv \rc \approx \ro (v_o/c_o)^{(q + 2)/(1-q)}.
 \label{rcrit} \eeq 
 
 The key physics which determines the nature of the cooling flow to
 accretion flow transition is whether $\rt \gsim \rc$ or $\rt \lsim
 \rc$.  That is, is the radius where the gravitational influence of
 the black hole becomes important inside or outside of the radius
 where, in the absence of the point mass, the cooling flow would have
 its sonic point?
 
 If $\rt \lsim \rc$ the cooling flow is oblivious to the existence of
 the point mass.  It undergoes its sonic transition with little
 modification because of the black hole.  Inside of the sonic point,
 the gas quickly cools towards zero temperature.  The reason is that
 in the supersonic zone a cooling flow in a logarithmic potential has
 $v \propto [\ln(r_s/r)]^{1/2}$ and $\rho \propto r^{-2} v^{-1} \sim
 r^{-2}$.  The velocity (density) therefore rises more slowly
 (quickly) with decreasing radius than in the subsonic zone, where the
 cooling time was of order the inflow time.  Consequently, in the
 supersonic zone the cooling time becomes much shorter than the inflow
 time and the gas cools rapidly (see, e.g., Fig. 4 of Sarazin \& White
 1987).  In this case, a simple cooling flow to accretion flow
 transition is not possible.  Whether such a cooling flow is an
 important source of mass for an accretion flow depends critically on
 the fate of the rapidly cooling matter inside the sonic point (e.g.,
 does it form stars which lose mass to feed the central black hole?).
 
 For M87, $\rt \approx 0.15$ kpc; this is comparable to $\rc$ if $q =
 0$.  Many other observed cooling flows have $\md \ \gg 10 \ \mpy$
 (M87's value) and would have $\rc \gg \rt$ if $q$ were equal to $0$.
 Significant mass drop out is, however, generally inferred in such
 cooling flows (for M87, $q \approx 0.6$ is appropriate; see below).
 From equation (\ref{scaling}) it follows that if $q \sim 1$ the Mach
 number of the cooling flow is roughly constant.  This insures that
 the gas remains subsonic down to the radius where the potential of
 the black hole begins to dominate the potential of the galaxy (i.e.,
 $\rt \gsim \rc$).  We now turn our attention to this regime.
 
 There is one conceptual complication in understanding the cooling
 flow to accretion flow structure in the regime where $\rt \gsim \rc$.
 Namely, in the point mass potential (i.e., for $r \ll \rt$), there
 are two viable analytical solutions to equations
 (\ref{mdot})-(\ref{energy}).  One is Bondi's adiabatic spherical
 accretion solution (Bondi 1952).  The other (valid only for $\gamma >
 11/7$) is a cooling flow solution (cf Fabian \& Nulsen 1977) which
 has $c_s^2 \propto r^{-1}$, $\rho \propto r^{-7/4}$ and $v \propto
 r^{-1/4}$.  It is not {\em a priori} obvious which of these solutions
 the cooling flow at large radii (in the galaxy's logarithmic
 potential) will match onto at small radii.  One might in fact
 suspect, on the basis of ``continuity'' arguments, that the cooling
 flow at large radii would match onto a cooling flow at small radii.
 
 Our calculations indicate, however, that this is not so.  The reason
 is simply that there is no {\em transonic} cooling flow solution in
 the $r^{-1}$ potential.  The Mach number of the cooling flow solution
 in a point mass potential decreases with decreasing radius as
 $r^{1/4}$; this solution cannot be matched to a sonic transition even
 with the introduction of a singularity into the point mass problem
 via the $1/(r-r_g)$ potential.\footnote{Since the Mach number of the
   cooling flow solution in the $1/r$ potential decreases with
   decreasing radius the only way it could ever make a sonic
   transition is if it were forced to by a singularity in the
   potential (or by some other boundary condition at small radii).}
 We have checked this carefully by numerically searching the $\md$ and
 $r_s$ space; we find that there is no transonic cooling flow solution
 for $r < \rt$.

 
 The upshot of this analysis is that the cooling flow makes a smooth
 transition to a nearly adiabatic Bondi flow in the vicinity of $\rt$.
 This provides a direct dynamical link between the cooling flow and
 the accretion flow onto the central black hole.
 
 
 The details of the flow structure near $\rt$ are shown in Figure 1,
 where we plot the temperature, density, Mach number, and the ratio of
 the cooling time to the inflow time for models with $q = 0.2,\, 0.6$,
 and $1$.  Figure 2 shows the accretion rate as a function of radius
 for these models.  The boundary conditions at $\ro = 100$ kpc are
 those appropriate for M87.  In M87, $\dot M$ is inferred to decrease
 from $\approx \, 10 \, \mpy$ at $\approx 70$ kpc to $\lsim \, 1 \,
 \mpy$ at a few kilo-parsecs (Stewart et al. 1984).  This is
 reasonably well captured by our $q = 0.6$ model.  The additional
 values of $q$ are shown to indicate the range of expected behavior.
 Note that the models in Figures 1 and 2 assume that $r_b = 0$,
 i.e., that the galaxy's potential is logarithmic everywhere.
 Modifications due to alternative potentials are discussed in \S3.1.
 
 The conceptually most important result in Figure 1 is the ratio of
 the cooling time to the inflow time.  This is $\sim 1$ at large radii
 where the accreting matter is indeed a cooling flow but it increases
 rapidly at smaller radii as the flow undergoes the transition to
 nearly adiabatic (Bondi) accretion.  Associated with this transition
 is a decrease in the importance of mass drop out.  In the Bondi
 regime, $\tc \gg \ti$ and so overdense regions do not have time to
 condense out of the flow.  This accounts for the nearly constant
 accretion rate for $r \lsim 300$ pc in Figure 2.

\subsection{The Importance of the Potential}

To illustrate the impact of the underlying potential of the host
galaxy on the structure of the cooling flow, consider the analytical
solution to equations (\ref{mdot})-(\ref{energy}) in the region where
the galaxy's mass profile flattens (i.e., for $\rt \lsim r \lsim r_b$
and $\beta < 1$): $c_s^2 \propto r^{-\beta + 1}$, $\rho \propto
r^{-5/4 - \beta/4}$, and $v \propto r^{-3/4 + \beta/4}$ (for
simplicity, we have taken $q = 0$).  Because the cooling flow is
roughly virial, the gas temperature decreases inward.  The Mach number
of the resulting solution therefore increases rapidly inwards, as
$r^{3\beta/4 - 5/4}$.  For observed values of $\beta \approx 0.25$,
the Mach number increases roughly as $r^{-1}$, much more rapidly than
the $r^{-1/2}$ scaling in the logarithmic potential.  One might worry
that this would change the results of the previous section by forcing
the cooling flow to have its sonic point outside of $\rt$.  Our
calculations show, however, that this does not happen.

Figure 3 shows the temperature, density, Mach number and ratio of
cooling time to inflow time for three models of M87 with $q = 0.6$,
and Figure 4 shows the corresponding predicted X-ray surface
brightness profiles.  The models in Figures 3 and 4 assume three
different forms for the potential of the host galaxy: one purely
logarithmic (solid line; the same model is also shown by a solid line
in Figs. 1 \& 2) and two with flattening mass profiles; $\beta \approx
0.25$ inside $r_b \approx 0.7$ kpc (dotted line) or $r_b \approx 3$
kpc (dashed-dot line).\footnote{M87's surface brightness profile is
  observed to flatten around $r_b \approx 0.7$ kpc (Faber et al.
  1995).}

As illustrated by Figures 3 and 4, the quantitative details of the
flow structure depend (not surprisingly) on the form of the potential
on kilo-parsec scales.  The most important result of this subsection,
however, is that even when the mass profile flattens, the cooling flow
still has its sonic point well inside $\rt$ and therefore still makes
a smooth transition to a nearly adiabatic accretion flow.  The
qualitative picture outlined in the previous section is therefore
unchanged.  This is partially because the inferred Mach number in M87
at a few kilo-parsec is sufficiently small ($\approx 0.04$; see Fig.
1c) that even if the Mach number increases linearly at small radii the
black hole still becomes important before the cooling flow has its
sonic point.  In addition, because the temperature of the gas
decreases somewhat if the galaxy's potential flattens, the black hole
becomes important for the dynamics of the gas at larger radii (the
transition radius, $\rt$, is effectively $\propto c_s^{-2}$ when $c_s
< \sigma$).  Note, for example, that $t_{\rm cool}/t_{\rm inf}$ is
actually larger at smaller radii in the model with $r_b = 3$ kpc.
This is because the decreasing temperature due to the flat mass
profile implies that the flow undergoes its transition from a cooling
flow to an accretion flow at a larger radius.


\section{Discussion}

The theoretical picture which emerges from the above analysis is a
very intuitive one: cooling flows in elliptical galaxies remain
subsonic down to roughly kilo-parsec scales at which point they are
gravitationally ``captured'' by the central supermassive black hole.
They then undergo a transition to a nearly adiabatic Bondi flow and
accrete onto the black hole.  There is consequently a direct and
simple dynamical transition from a cooling flow to an accretion flow.


Our calculations indicate that only if the cooling flow accretion rate
on scales of a few kilo-parsecs is $\gsim \ 10-100 \ \mpy$ will the
cooling flow have its sonic transition before being captured by the
black hole (the precise $\dot M$ depends on the black hole mass --
assumed $\gsim 10^9 M_\odot$ -- and the galaxy potential, i.e.,
$\beta$, $\sigma$, and $r_b$; less massive black holes, larger
$\sigma$, and flatter galaxy mass profiles -- larger $r_b$ and smaller
$\beta$ -- require smaller $\dot M$).  In this case, the gas cools
rapidly just inside its sonic point, which lies outside the region
where the gravitational influence of the black hole is important.
Although $\dot M \gsim 10-100 \ \mpy$ is readily achieved on $\sim
100$ kpc scales, mass drop out due to thermal instability, and the
consequent decrease of $\dot M$ with decreasing radius, implies that
observed cooling flows typically lie in the parameter regime where a
smooth transition from a cooling flow to a nearly adiabatic accretion
flow is the viable transonic solution.  We find that this conclusion
is valid even given uncertainties in the mass profile of the host
galaxy on kilo-parsec scales (\S 3.1).

It is interesting to note that all of the solutions shown in Figures
1-5 have $r_s \ll \rt$; that is, the sonic point lies well inside the
radius where the potential changes from that of the galaxy to that of
the black hole.  Nonetheless, the dynamics of the accreting gas
changes from that of a cooling flow to that of a Bondi flow at, or
even outside of, $\rt$.  The requirement that the flow undergo a
smooth sonic transition influences its structure out to radii $\gsim
\rt \gg r_s$.


The primary assumption of our analysis is that there is negligible
angular momentum in the inflowing material.  Angular momentum may,
however, become important on the small scales of interest here; if so,
a transition to a Bondi flow will not be viable.  In this case, we
believe that the cooling flow will generally undergo a transition to
an advection-dominated accretion flow (ADAF), the hot accretion flow
analogue of Bondi's solution for flows with angular momentum (Narayan
\& Yi 1994, 1995; Abramowicz et al. 1995).  It is well known that
ADAFs can only exist below a critical accretion rate, $\dot M_c$.  In
the present context, this can be estimated by requiring the cooling
time at $\rt$ to be greater than the inflow time, which yields \beq
\dot M_c \sim 10 \ \alpha^2 \ \mpy \ \left(M \over 10^9 M_\odot\right)
\left(\sigma \over 300 \ \kms\right).
\label{mdc} \eeq
In equation (\ref{mdc}), $\alpha$ is the ratio of the inflow velocity
to the local Keplerian speed, and is roughly the Shakura \& Sunyaev
(1973) viscosity parameter.  

The properties of the transition from a cooling flow to a Bondi flow
described in \S3 are determined primarily by the low radiative
efficiency of the Bondi solution.  Consequently, the qualitative
picture outlined in this paper should apply equally well to the
transition from a cooling flow to an ADAF, since the latter is also
radiatively inefficient (the quantitative details may, of course,
change).  If, however, the accretion rate supplied by the cooling flow
to $\rt$ is $\gsim \dot M_c$, the cooling flow may not be able to
evolve into an ADAF; it will instead likely collapse to a thin disk.
We plan to explore this question in a subsequent paper.




\subsection{Observational Prospects}

To conclude, we discuss the observational prospects for observing the
transition from a cooling flow to an accretion flow with the excellent
($\lsim 1$ arcsec) angular resolution of the \cxo.  We focus our
discussion on M87 in the Virgo cluster, as it is the closest large
cooling flow.  For a distance to M87 of $\approx 18$ Mpc, one arcsec
is $\approx 0.1$ kpc.  The basic inference from Figures 1 \& 3 is that
on scales of $0.1$ kpc, the dynamics of accreting gas in M87 is far
from that of a cooling flow.  The ratio of the cooling time to the
inflow time is likely $\approx 10$ or larger.

One prediction of our model is that the inferred gas temperature in
M87 should increase with decreasing radius on sufficiently small
scales (when the black hole's potential dominates).  In particular,
all of our models show an increasing temperature profile for $r \lsim
300$ pc.  Note however, that because the gas temperature in nearly all
models is roughly virial (and therefore traces the potential), such a
temperature increase would not necessarily confirm the cooling flow to
accretion flow transition.  It would merely show that the gas dynamics
is dominated by a centrally condensed mass distribution (namely, the
black hole).

The second prediction of our model is that the X-ray surface
brightness profile, while centrally peaked, should be suppressed with
respect to that expected from a cooling flow without a central point
mass.  This is because the transition from a cooling flow to an
accretion flow entails a decrease in radiative efficiency.  A
suppressed X-ray surface brightness profile is, however, also the
characteristic signature of a cooling flow with mass drop out. 

At a minimum, our analysis indicates that significant care must be
taken in interpreting the X-ray surface brightness profiles of
elliptical galaxies on $\lsim 1$ kpc scales.  Standard analysis
techniques, based on the assumption of a cooling flow, could
incorrectly infer a radially decreasing accretion rate when in fact
the accretion is via a Bondi flow with constant $\dot M$ and it is the
radiative efficiency that decreases.  As shown in \S3 (see Figs. 2 \&
5), $\dot M$ is expected to be nearly constant on the small scales of
interest here since the transition to a Bondi-like flow strongly
suppresses mass drop out due to thermal instability.

Given an observed X-ray luminosity $L_X$ and sound speed $c_s$, the
true local accretion of the flow is \beq \dot M \approx {L_X \over
  c_s^2} {t_{\rm cool} \over t_{\rm inf}}. \label{mdot2}\eeq Our
calculations indicate that $t_{\rm cool}/t_{\rm inf}$ is likely to be
$\gsim 10$ on $\approx 0.1$ kpc scales in M87 so that accretion rates
inferred assuming a cooling flow ($\tc \approx \ti$, $\dot M \approx
L_X/c^2_s$) can be underestimates by up to an order of magnitude.
This is shown explicitly in Figure 5 where we compare the true
accretion rate for our baseline ($q = 0.6$) model of M87 (solid line)
with that which we would infer if we took the predicted X-ray surface
brightness and temperature profiles and assumed a pure cooling flow
(dotted line).

X-ray spectra may break the degeneracy between a change in radiative
efficiency and mass drop out.  The presence of co-spatial
multi-temperature components in coolings flows is indicated by a
detailed examination of their X-ray spectral lines (Canizares et al.
1979, 1982; Mushotzky et al.  1981).  This is interpreted as evidence
for mass drop out in the cooling flow.  Consequently, the {\em
  absence} of evidence for such multi-temperature components (or a
decrease in their prominence), together with a suppression in the
X-ray surface brightness at small radii, would provide considerable
support for our model of the cooling flow to accretion flow
transition.

\acknowledgements We acknowledge support from an NSF Graduate Research
Fellowship (EQ) and NSF Grant 9820686 (RN).  We thank the referee for
useful comments which significantly improved this paper.



\newpage
 
\begin{figure}
\plotone{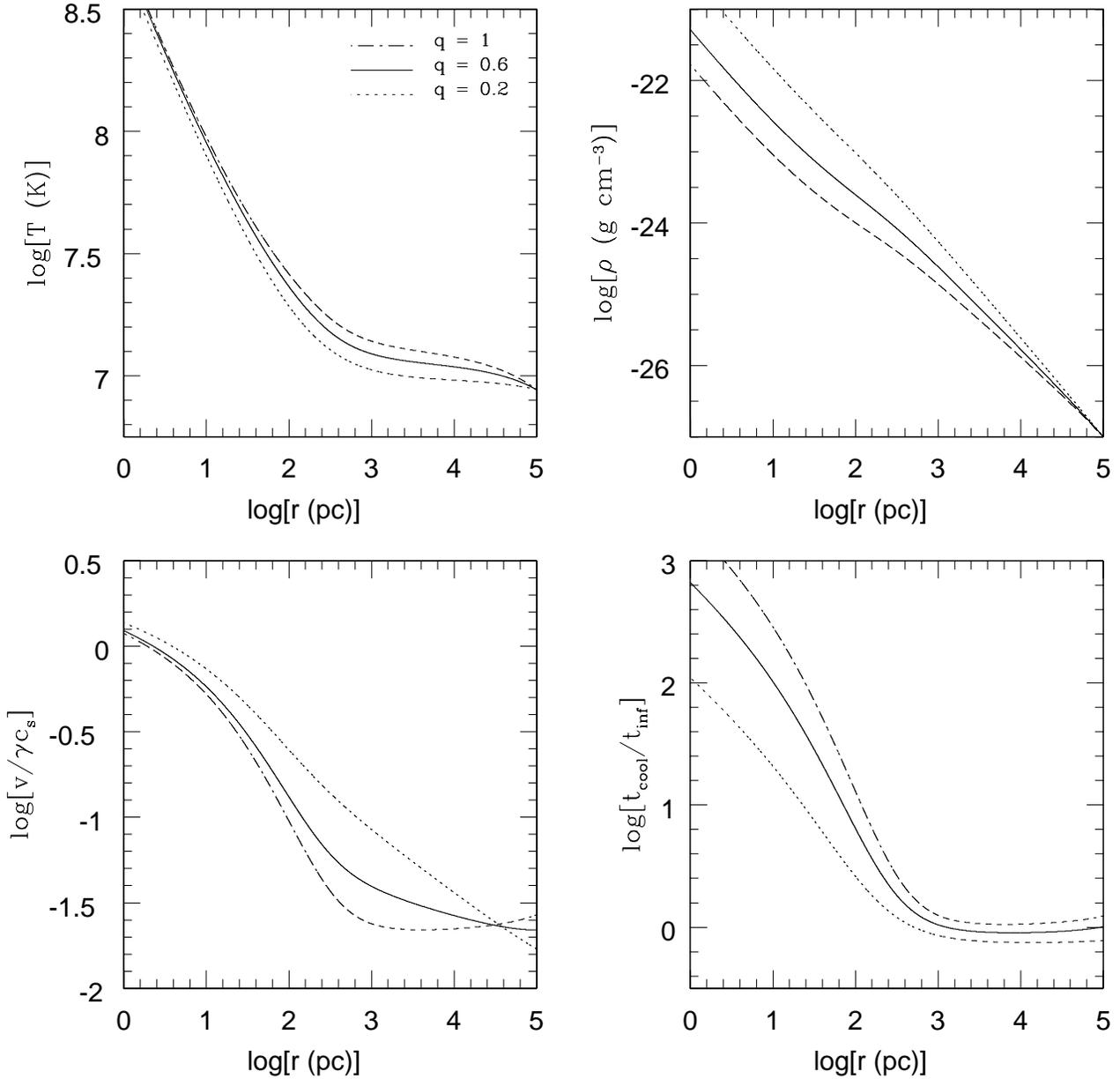}
\caption{Numerical models of the transition from a cooling flow to an 
  accretion flow.  The temperature, density, ratio of cooling time to
  inflow time, and Mach number are shown (clockwise, from top left).
  The outer boundary conditions are chosen to represent M87. The $q =
  0.6$ model reproduces M87's radially varying $\dot M$ reasonably
  well (Fig. 2). Models for two additional values of $q$ are shown to
  indicate the range of expected behavior.}
\end{figure}

\newpage
\begin{figure}
  \plotone{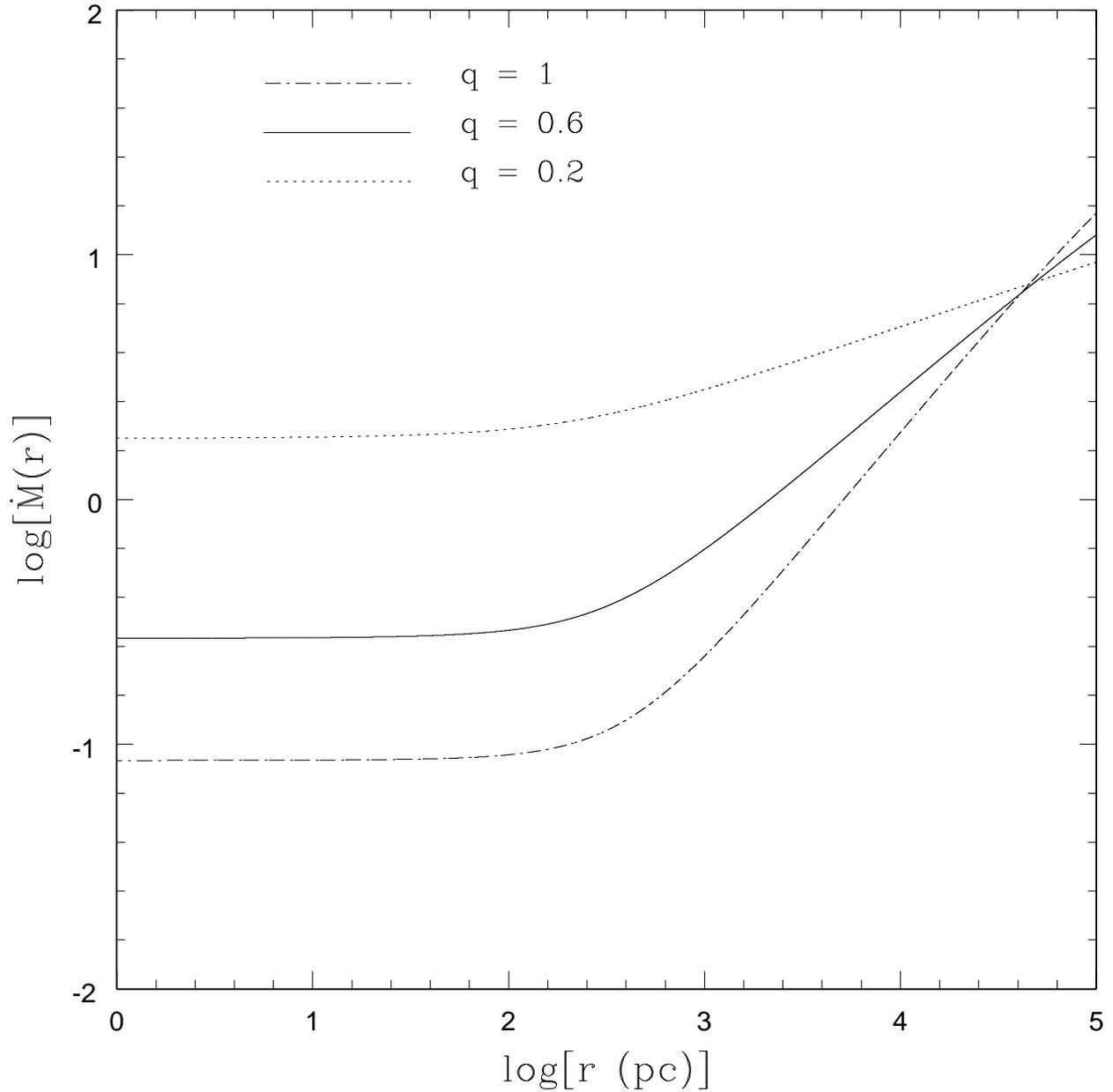} \caption{The accretion rate as a function of
    radius for the three models whose structure is shown in Figure 1.
    The $q = 0.6$ model roughly represents M87, for which observations
    show a decrease in $\dot M$ from $\approx 10 \, \mpy$ at $\approx
    70$ kpc to $< 1 \, \mpy$ at a few kilo-parsecs.  The presence of a
    central black hole (taken to have $M = 3 \times 10^9$) and the
    associated transition to a nearly adiabatic accretion flow leads
    to a complete suppression of mass drop out on small scales.  This
    accounts for the constant accretion rate on scales less than
    $\approx 300$ parsecs.}
\end{figure} 
\newpage
\begin{figure}
\plotone{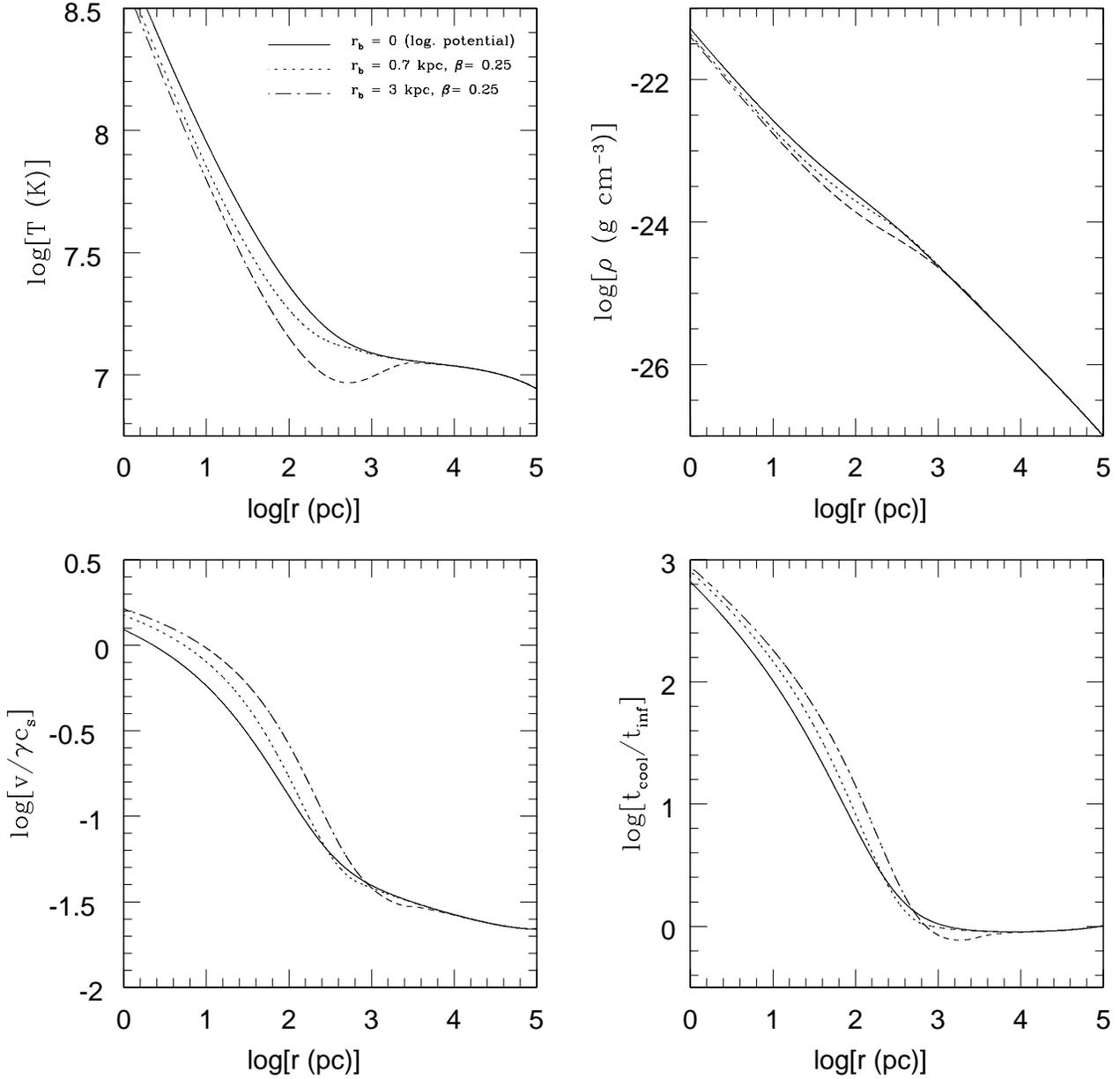}
\caption{Numerical models of the cooling flow to accretion flow transition
  for three possible underlying potentials for the host galaxy (see
  eqs. [\ref{fg1}] and [\ref{fg2}]).  The outer boundary conditions
  are chosen to represent M87.  The temperature, density, ratio of
  cooling time to inflow time, and Mach number are shown (clockwise,
  from top left).}
\end{figure}
\newpage
\begin{figure}
  \plotone{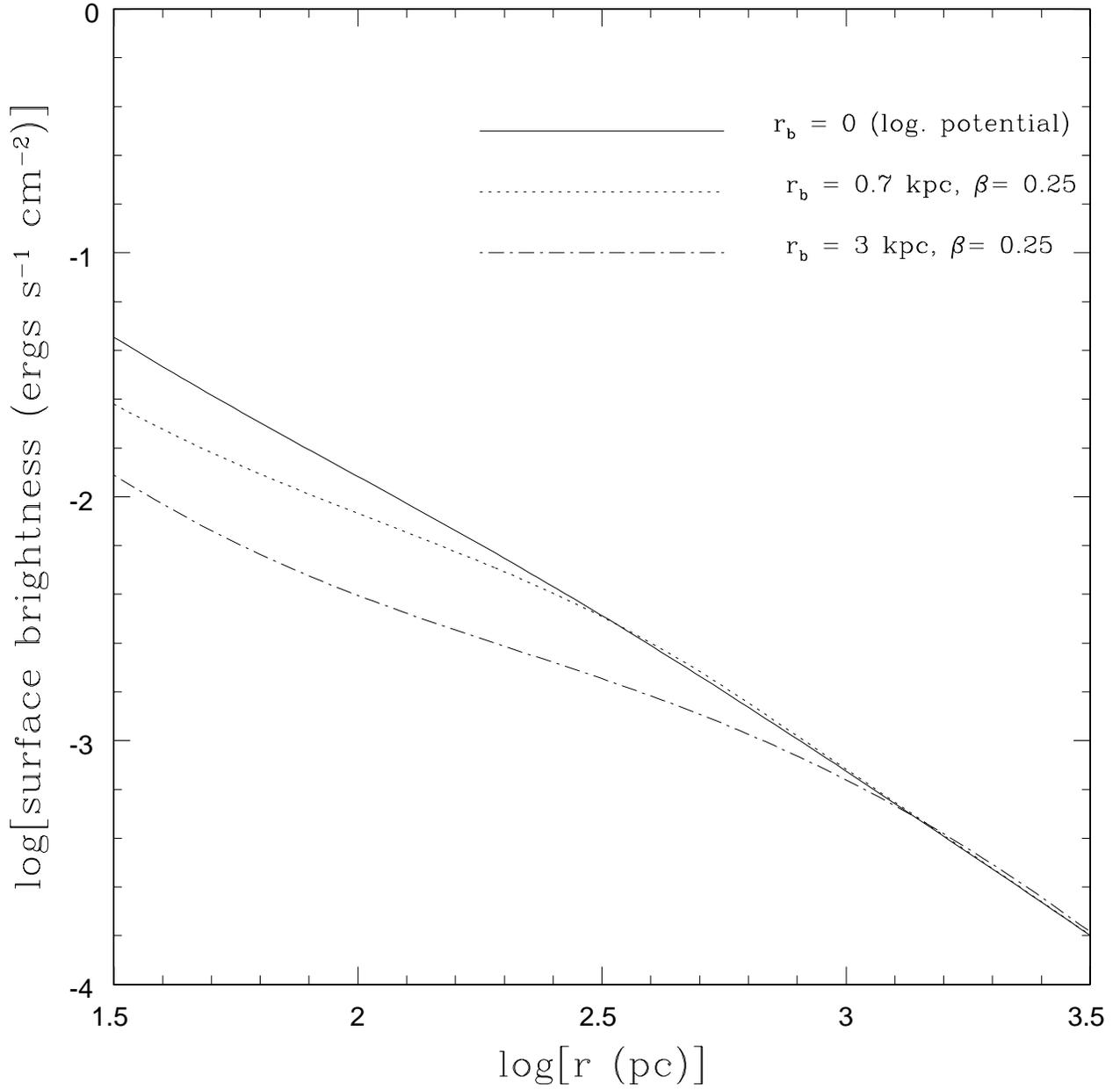} \caption{Predicted X-ray surface brightness
    profiles for the three models whose structure is shown in Figure
    3.}
\end{figure} 
\newpage
\begin{figure}
  \plotone{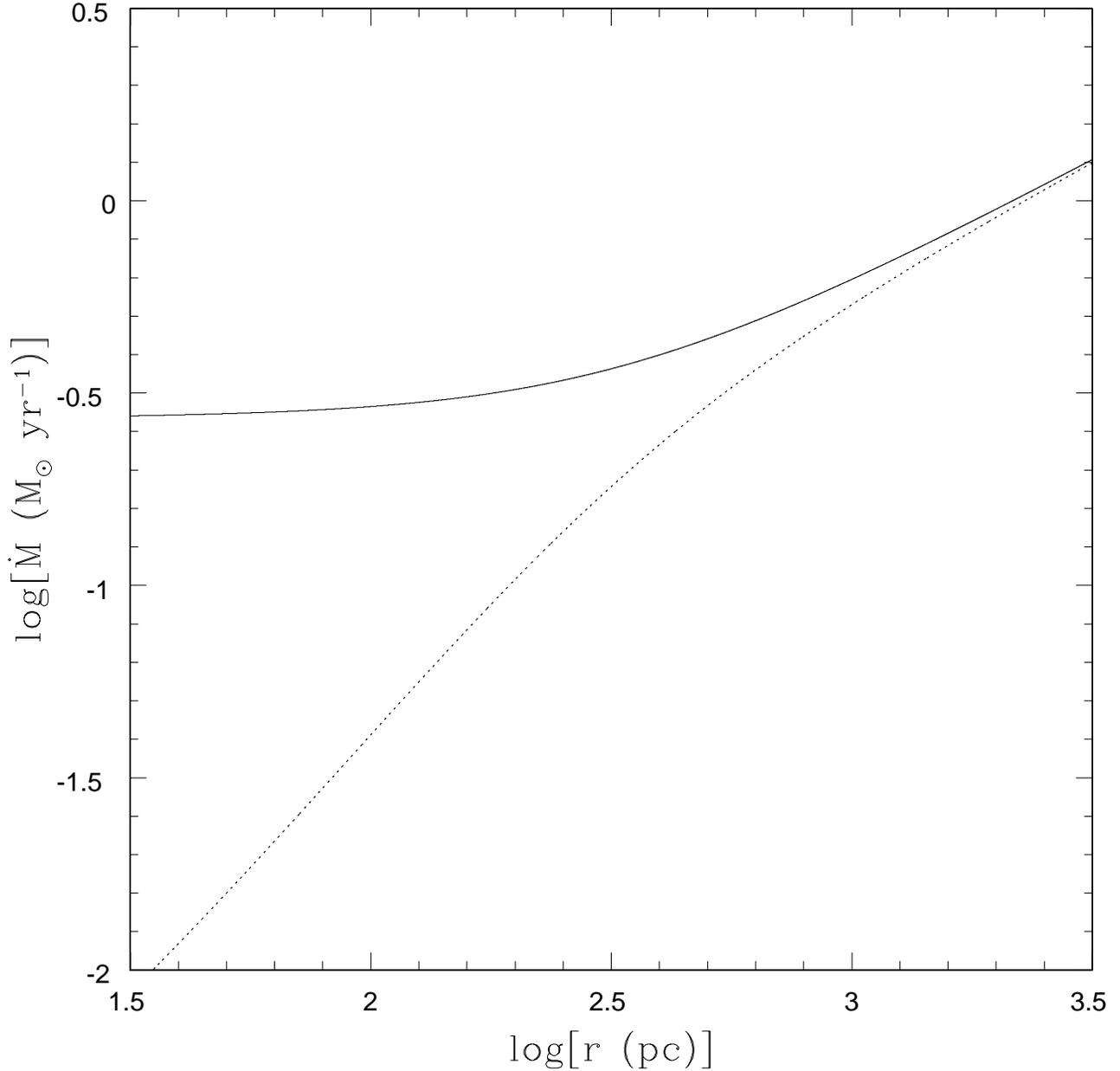} \caption{The solid line shows the true radially
    varying accretion rate for our $q = 0.6$ model of M87 (pure
    logarithmic potential; solid line in Figs. 3 \& 4).  The dotted
    line shows the accretion rate inferred from the predicted X-ray
    surface brightness and temperature profiles if one assumes that
    the gas is a pure cooling flow. The true accretion rate is nearly
    constant on the small scales shown in this figure because the
    transition from a cooling flow to a nearly adiabatic accretion
    flow suppresses mass drop out due to thermal instability.  The
    decreasing radiative efficiency associated with this transition
    could, however, be misinterpreted as a radially decreasing
    accretion rate (dotted line).}
\end{figure} 

\end{document}